\documentclass[twocolumn,showpacs,aps,prl,superscriptaddress]{revtex4}
\usepackage{graphicx}
\usepackage{dcolumn}
\usepackage{epsfig}
\usepackage{amsmath}
\RequirePackage{xspace}
\newcommand{\BaBarYear}    {06}
\newcommand{\BaBarNumber}  {071}
\newcommand{\SLACPubNumber} {12257}
\def\epem  {\ensuremath{e^+e^-}\xspace}

\newcommand{\xf}{\mbox{${\cal F}$}}
\def\KS    {\ensuremath{K^0_{\scriptscriptstyle S}}}
\def\Bu      {\ensuremath{B^+}}
\def\Bub     {\ensuremath{B^-}}
\def\Bbar    {\overline{B}{}}

\def\Bzb     {\ensuremath{\Bbar^0}}
\def\Bz      {\ensuremath{B^0}}
\def\BpBm    {\ensuremath{\Bu  \Bub}}
\def\BzBzb   {\ensuremath{\Bz  \Bzb}}
\newcommand{\DE}{\ensuremath{\Delta E}}
\newcommand{\pvec}{{\bf p}}
\newcommand{\half}{\mbox{${1\over2}$}}
 \def\mes{\mbox{$m_{\rm ES}$}}
 \def\angular{\mbox{$\cal H$}}
\newcommand{\calB}{\mbox{${\cal B}$}}

\newcommand{\aunob}{\mbox{$a_1$}}
\newcommand{\aunobpi}{\mbox{$a_1 \pi$}}

\newcommand{\Nappim}{\mbox{$a^{\pm}_1(1260)\, \pi^{\mp}$}}

\newcommand{\Nadueppim}{\mbox{$a^{\pm}_2(1320)\, \pi^{\mp}$}}

\newcommand{\Nbtoappim}{\mbox{$B^0 \rightarrow a^{\pm}_1(1260)\, \pi^{\mp}  $}}
\newcommand{\NNbtoappim}{\mbox{$B^0 \rightarrow a^{\pm}_1\, \pi^{\mp}  $}}
\newcommand{\Nbtoadueppim}{\mbox{$B^0 \rightarrow a^{\pm}_2(1320)\, \pi^{\mp}  $}}

\newcommand{\Natopipipi}{\mbox{$ a^{\pm}_1(1260) \rightarrow \pi^{\mp}\pi^{\pm}\pi^{\pm}  $}}

\newcommand{\UfourS}{\mbox{$\Upsilon(4S)$}}

\def\babar{{\em B}{\footnotesize\em A}{\em B}{\footnotesize\em AR}}
\def\BB{\mbox{$B\overline B\ $}}
\def\pep2{PEP-II}

\newcommand\etal{{\it et al.}}
\newcommand{\dedx}{\ensuremath{\mathrm{d}\hspace{-0.1em}E/\mathrm{d}x}}

\newcommand{\gev}{\mbox{$\textrm{GeV}$}} 
\newcommand{\mev}{\mbox{$\textrm{MeV}$}} 
 
\newcommand{\bflav}{\ensuremath{B_{\rm flav}}}

\providecommand{\dt}{\deltat}
\providecommand{\sigdt}{\ensuremath{\sigma_{\deltat}}}

\newcommand{\ttag}{\ensuremath{t_{\rm tag}}}

\def\qqbar{\mbox{$q\bar q\ $}}

\def\a1pi{\aunob\pi}

\def\CP                {\ensuremath{C\!P}\xspace}
\def\ra                 {\ensuremath{\rightarrow}\xspace}
\def\to                 {\ensuremath{\rightarrow}\xspace}

\def\deltat {\ensuremath{\Delta t}\xspace}
\newcommand{\beqn}{\begin{eqnarray}}
\newcommand{\eeqn}{\end{eqnarray}}
\def\deltamd{\ensuremath{{\rm \Delta}m_d}\xspace}

\def\Ampbar{\kern 0.18em\overline{\kern -0.18em A}{}_{{a_1}\pi}}
\def\Acpapi{{\cal A}_{CP}^{{a_1}\pi}}

\def\dC{\ensuremath{\Delta C}\xspace}
\def\dS{\ensuremath{\Delta S}\xspace}

\def\dCa1pi{\ensuremath{\Delta C}_{{a_1}\pi}}
\def\dSa1pi{\ensuremath{\Delta S}_{{a_1}\pi}}

\def\Ca1pi{\ensuremath{C_{{a_1}\pi}}}
\def\Sa1pi{\ensuremath{S_{{a_1}\pi}}}
\def\Acpa1pi{A_{CP}^{{a_1}\pi}}

\newcommand{\jprlBase}  [1]     {Phys.\ Rev.\ Lett. \xspace}
\newcommand{\jprl}      [1]    {\jprlBase\ {\bf #1}}
\newcommand{\jprBase}        {Phys.\ Rev.\ \xspace}
\newcommand{\jprd}      [1]  {\jprBase\ D~{\bf #1}}
\newcommand{\plBase}   [1]         {Phys.\ Lett.\xspace}
\newcommand{\plb}      [1]    {\plBase\ B~{\bf #1}}
\newcommand{\nimBaseA}       {Nucl.\ Instr.\ Meth.\ \xspace }
\newcommand{\nima}      [1]  {\nimBaseA~A~{\bf #1}}
\newcommand{\zpBase}         {Z.\ Phys.}
\newcommand{\zpc}       [1]  {\zpBase\ C~{\bf #1}}
\newcommand{\npBase}         {Nucl.\ Phys.\ \xspace}
\newcommand{\npb}       [1]  {\npBase\ B~{\bf #1}}
\newcommand{\jpg}       [1]  {{J.\ Phys.\ {\bf G{\bf #1}}}}
\newcommand{\progtp}    [1]  {{Prog.\ Theor.\ Phys.\ {\bf #1}}}

\newcommand{\jmplBase}  [1]     {Mod.\ Phys.\ Lett.}
\newcommand{\jmpl}      [1]    {\jmplBase\ A~{\bf #1}}

\newcommand{\epjBase}  [1]     {Eur.\ Phys.\ J. \xspace}
\newcommand{\epj}      [1]    {\epjBase\ C~{\bf #1}}

\def\figurebox#1#2#3{%
    \def\arg{#3}%
    \ifx\arg\empty
    {\hfill\vbox{\hsize#2\hrule\hbox to #2{\vrule\hfill\vbox to #1{\hsize#2\vfill}\vrule}\hrule}\hfill}%
    \else
    {\hfill\epsfbox{#3}\hfill}%
    \fi}

\begin{document}

\begin{flushleft}
\babar-PUB-\BaBarYear/\BaBarNumber \\
SLAC-PUB-\SLACPubNumber\\
hep-ex/0612050
\end{flushleft}

\title{\large  \bf\boldmath  Measurements of \CP-Violating Asymmetries in $B^0 \ra \Nappim $ Decays}

%
\author{B.~Aubert}
\author{M.~Bona}
\author{D.~Boutigny}
\author{Y.~Karyotakis}
\author{J.~P.~Lees}
\author{V.~Poireau}
\author{X.~Prudent}
\author{V.~Tisserand}
\author{A.~Zghiche}
\affiliation{Laboratoire de Physique des Particules, IN2P3/CNRS et Universit\'e de Savoie, F-74941 Annecy-Le-Vieux, France }
\author{E.~Grauges}
\affiliation{Universitat de Barcelona, Facultat de Fisica, Departament ECM, E-08028 Barcelona, Spain }
% ...
\collaboration{The \babar\ Collaboration}
\noaffiliation

\begin{abstract}
\noindent
We present measurements  of \CP-violating asymmetries in the decay 
\Nbtoappim\  with \Natopipipi. The data sample corresponds to $384
\times 10^6$ \BB\ pairs  collected with the \babar\ detector at the \pep2
asymmetric $B$-factory at SLAC. We measure the \CP-violating asymmetry
$\Acpapi = -0.07 \pm 0.07 \pm 0.02$, 
 the mixing-induced \CP violation parameter  $\Sa1pi = 0.37 \pm 0.21\pm 0.07$, the direct \CP violation parameter $\Ca1pi = -0.10 \pm 0.15\pm 0.09$, and the parameters $\dCa1pi = 0.26 \pm 0.15\pm 0.07$ and $\dSa1pi = -0.14 \pm 0.21 \pm 0.06$. From these measured quantities we determine the angle $\alpha_{\rm eff} = 78.6^{\circ} \pm 7.3^{\circ}$.
\end{abstract}


\maketitle
The angle $\alpha \equiv \arg\left[-V_{td}^{}V_{tb}^{*}/V_{ud}^{}V_{ub}^{*}\right]$
of the unitarity triangle of the Cabibbo-Kobayashi-Maskawa (CKM) quark-mixing
 matrix \cite{CKM} has recently been measured
by the \babar\ and Belle Collaborations 
from time-dependent \CP asymmetries in the $B^0$ decays to $\pi^+\pi^-$ \cite{pipi}, 
$\rho^{\pm}\pi^{\mp}$ \cite{rhopi}, and $\rho^+ \rho^-$ \cite{rhorho}.
The decay  $B^0$ to $\a1pi$ \cite{a1} proceeds dominantly through the 
  $\bar{b} \ra \bar{u} u \bar{d}$    process in the same way as the 
     previously studied modes.  However, due to the presence of
     additional loop contributions, these measurements determine an
     effective value $\alpha_{\rm eff}$, rather than $\alpha$ itself.
 This obstacle can be overcome using isospin symmetry \cite{GL}, with bounds to $\Delta \alpha =\alpha - \alpha_{\rm eff}$
determined using either an isospin analysis \cite{Gross} or broken  SU(3) flavor symmetry \cite{GZ}.
Because it has the smallest contribution from loop diagrams, the $B^0
\ra\rho^+ \rho^-$  decay currently allows the most precise single determination of $\alpha$  \cite{Adrian}.
The  \babar\  collaboration recently reported
the observation of $B^0 \ra a_1^{\pm}\pi^{\mp}$ \cite{a1pi},
where the angle $\alpha_{\rm eff}$ can  be determined by measuring
time-dependent \CP asymmetries \cite{Aleksan,Zupan}.
The state $a_1^{\pm}\pi^{\mp}$, like  $\rho^{\pm}\pi^{\mp}$,  is not a \CP
 eigenstate and four flavor-charge configurations must be considered
 $(\Bz(\Bzb) \ra  a_1^{\pm}\pi^{\mp}$).
Theoretical bounds on  $\Delta \alpha$ in these decay modes based on SU(3) flavor symmetry have been derived in Ref. \cite{Zupan}. 
 
In this Letter we report measurements of the \CP  parameters in the 
decay \NNbtoappim\ with  $a_1^{\pm} \ra \pi^{\mp} \pi^{\pm}
\pi^{\pm}$. 
The analysis is done in 
the quasi-two-body approximation \cite{SLAC}.
The data were collected with the \babar\ detector~\cite{BABARNIM}
at the PEP-II asymmetric $e^+e^-$ collider~\cite{pep}. An integrated
luminosity of 349~fb$^{-1}$, corresponding to
384 $\pm$ 4 million \BB\ pairs, was recorded near the $\Upsilon (4S)$ resonance
(``on-resonance'') at a  center-of-mass (CM) energy $\sqrt{s}=10.58~\gev$.
An additional 37~fb$^{-1}$ were taken about 40~MeV below
this energy (``off-resonance'') for the study of continuum background in
which a charm or lighter quark pair is produced.

From a candidate \BB\ pair we reconstruct a \Bz\  decaying into the final 
state $f= \aunobpi$ ($B^0_{\aunobpi}$).  We also reconstruct the vertex of 
the other $B$ meson ($B^0_{\rm tag}$) and identify its flavor.
The difference $\deltat \equiv t_{\aunobpi} - \ttag$
of the proper decay times of the reconstructed and tag $B$ mesons, 
respectively, is obtained from the measured distance between the $B^0_{\aunobpi}$
and  $B^0_{\rm tag}$ decay vertices and from the boost ($\beta \gamma =0.56$) of 
the \epem system. The \deltat\ distributions are given \cite{Zupan} by:
\beqn
\label{eq:thTime}
  \lefteqn{F^{a_1^{\pm} \pi^{\mp}}_{Q_{\rm tag}}(\deltat) = (1\pm \Acpapi)
           \frac{e^{-\left|\deltat\right|/\tau}}{4\tau} \bigg\{ 1 - Q_{\rm
             tag} \Delta w +}\\
	&&\hspace{1.cm} Q_{\rm tag} (1-2w) 
             \bigg[(\Sa1pi \pm \dSa1pi)\sin(\deltamd\deltat)-\nonumber\\[-0.1cm]
	&&\hspace{1.cm}\phantom{Q_{\rm tag} (1-2w) \bigg[}
		(\Ca1pi\pm \dCa1pi)\cos(\deltamd\deltat)\bigg]\bigg\}\;,\nonumber
\eeqn
where $Q_{\rm tag}= 1(-1)$ when the tagging meson $\Bz_{\rm tag}$
is a $\Bz(\Bzb)$, $\tau$ is the mean 
\Bz\ lifetime, $\deltamd$ is the mass difference between the two $B^0$ 
mass eigenstates, and the mistag parameters $w$ and $\Delta w$ are the
average and difference, respectively, of the probabilities that a true
$\Bz$ is incorrectly tagged as a $\Bzb$ or vice versa.
The time- and flavor-integrated  charge asymmetry $\Acpapi$ measures 
direct \CP violation.
The quantities $\Sa1pi$ and $\Ca1pi$ 
parameterize the mixing-induced \CP violation related to the angle $\alpha$,
and flavor-dependent direct \CP violation, respectively.
The parameter $\dCa1pi$ describes the asymmetry between the rates 
$\Gamma({\Bz} \to{a_1^+\pi^-}) + \Gamma({\Bzb} \to {a_1^-\pi^+})$ and
${\Gamma(\Bz} \to {a_1^-\pi^+}) + \Gamma({\Bzb} \to {a_1^+\pi^-})$, while
$\dSa1pi$ is related to the strong phase difference between
the amplitudes contributing to $\Bz \to\a1pi$ decays. The parameters $\dCa1pi$ and $\dSa1pi$ are insensitive to
\CP violation.
The flavor-tagging algorithm
uses six mutually exclusive  categories. Its  analyzing 
power is measured to be $( 30.4\pm 0.3)\%$ \cite{s2b}. 

Charged particles are detected and their momenta measured by the
combination of a silicon vertex tracker (SVT), consisting of five layers
of double-sided silicon detectors, and a 40-layer central drift chamber,
both operating in the 1.5-T magnetic field of a superconducting solenoid.
Charged-particle identification (PID) is provided by the average
energy loss (\dedx) in the tracking devices and by an internally reflecting ring-imaging
Cherenkov detector (DIRC) covering the central region.
Separation between pions and kaons is achieved at the 
        level of four standard deviations ($\sigma$) for momenta below
        3~\gev, decreasing to 2.5 $\sigma$ at 4~\gev.

Full Monte Carlo (MC) simulations \cite{geant4} of the signal decay modes, 
continuum, and \BB\ backgrounds   are used to establish 
the event selection criteria.
The MC signal events are simulated as $B^0$ decays to $a_1 \pi$
with $a_1 \rightarrow \rho \pi$.
For  the  ${a_1}$ meson parameters we take the mass  
$m_0=1230$ \mev\ and the width $\Gamma_0=400$ \mev\ ~\cite{evtgen,PDG2006}. 

We reconstruct the decay $\aunob \ra 3 \pi$ with the
following requirement on the invariant mass: $0.87<m_{\aunob}<1.8$~\gev.
The intermediate dipion state is reconstructed
with an invariant mass between 0.51 and 1.1~\gev. 
We impose  several PID requirements to ensure the
identity of the signal pions. For the decay pion coming  from the $B$ meson
 we require the measured  Cherenkov angle to be within
$-2\,\sigma$ and $+5\,\sigma$ from the expected value for a pion.
This requirement removes 98.6\% of the background from
$\aunob K$. 
A $B$ candidate is characterized kinematically by the energy-substituted 
mass $\mes = \sqrt{(s/2 + \pvec_0\cdot \pvec_B)^2/E_0^2 - \pvec_B^2}$ and
energy difference $\DE = E_B^*-\half\sqrt{s}$, where the subscripts $0$ and
$B$ refer to the initial \UfourS\ and to the $B$ candidate in the
laboratory frame, respectively, 
and the asterisk denotes the CM frame. The resolutions in \mes\ and 
in \DE\ are  about 3.0  \mev\ and  20 \mev\ respectively.
We require $|\DE|\le0.1$ GeV and $5.25\le\mes\le5.29\ \gev$. To
reduce the number of false $B$-meson
candidates we require that the probability of the  $B$ vertex fit be
greater than 0.01. 
The absolute value of the cosine 
of the angle between the direction of the
$\pi$ meson from  $\aunob \ra \rho \pi$   with respect to the flight direction of the $B$ 
in the \aunob\ meson rest frame
is required to be less than $0.85$ to suppress combinatorial
background. The distribution of this variable is uniform for signal
and peaks near unity for this background.

To reject continuum background, we use
the angle $\theta_T$ between the thrust axis of the $B$ candidate and
that of the rest of the tracks and neutral clusters in the event, calculated in
the center-of-mass frame. The distribution of $\cos{\theta_T}$ is
sharply peaked near $\pm1$  for $q\bar q$ candidates, which
have a  jet-like topology, and is nearly uniform for the isotropic $B$-meson decays. We require
$|\cos{\theta_T}|<0.65$.
We discriminate further against \qqbar\ background with a
Fisher discriminant \xf\ that combines several variables that
characterize the production dynamics and energy flow in the event
\cite{Fisher}.   The remaining continuum background is modeled from off-resonance data. 

We use MC simulations of \BzBzb\ and \BpBm\ decays to look for \BB\ backgrounds, 
which can come from $B$ decays with or without charmed particles in
the final state.
Neutral and charged $D$ mesons may contribute to background  through particle mis-identification or 
mis-reconstruction. We remove any combinations of the decay products,
including possible additional $\pi^0$, with  invariant mass consistent with  nominal mass values for 
$D^{\pm}\ra K^{\mp} \pi^{\pm}\pi^{\pm}$ or $\KS\ \pi^{\pm}$ 
and  $D^0\ra K^{\mp} \pi^{\pm}$  or $K^{\mp}\pi^{\pm}\pi^0$. 
The decay mode \Nbtoadueppim\   has the same 
final-state particles as the signal. We suppress  this decay with the angular 
variable \angular\ 
, defined as  the cosine
of the angle between the normal to the plane of the $3\pi$ resonance
and the flight direction of the primary pion from $B$ meson evaluated in the $3\pi$
resonance rest frame. Since the  $a_1$ and  $a_2(1320)$ mesons have
spins of 1 and 2 respectively,  the distributions of 
the variable \angular\ for these two resonances differ. We require 
$|\angular|< 0.62$.

The average number of candidates found per selected event is 1.32. 
In the case of events with multiple candidates we choose the candidate 
with the best $B$-vertex fit probability.  From simulated signal events we find that this algorithm selects 
the correct candidate in about 92\% of the events containing multiple 
candidates, and introduces negligible bias.

We obtain the \CP  parameters and signal yield from an unbinned extended  maximum 
likelihood (ML) fit with the input observables \DE, \mes, \xf,
$m_{\aunob}$, \angular, and \deltat.  
We have six fit components in the likelihood: signal, charm and 
charmless \BB\ background, 
 \Nbtoadueppim, continuum \qqbar\ background, and non-resonant $\rho \pi \pi
 $. The charmless component also includes candidates that
 were incorrectly reconstructed from particles in events that contain
 a true signal candidate.
Based on measurements of branching fractions for similar charmless
decays, we assume   $\calB(B^0\ra\rho^0\pi^+\pi^-)=(2 \pm2) \times 10^{-6}$,
which corresponds to 19 expected events in the ML fit sample. This
yield is fixed in the fit and a systematic uncertainty is assigned  to the final results.

The total probability density function (PDF) for the component $j$ and 
tagging category  $c$ in the event $i$,  ${\cal P}_{j,c}^i$,  is written as a product 
of the PDFs of the discriminating variables used in the fit.
The factored form of the PDF is a good approximation
since linear correlations among observables are below 10\%.
The systematic uncertainty from residual correlations is taken into
account in the fit bias.
We write the extended likelihood function for all events as
\begin{equation}
{{\cal L}} =  \prod_c  \exp{(-n_c)} \prod_i^{N_c} \left[ \sum_{j}n_j
f_{j,c}  {\cal P}_{j,c}^i  \right]\,,
\end{equation}
where $n_j$ is the yield of events of component $j$,  $f_{j,c}$ is the fraction of events
of component $j$ for each category $c$, 
$n_c = \sum_j f_{j,c}n_j $  is the number of events found by the fitter for category $c$, and $N_c$ is 
the number of events of category $c$ in the sample.  
We fix $f_{j,c}$  to $f_{\bflav,c}$, the values measured with a
large
sample of fully reconstructed $B^0$ decays into flavor eigenstates
(\bflav\ sample)   \cite{Resol}, for the  signal, $\rho\pi\pi$, and  \Nbtoadueppim\
fit components.  We fix $f_{j,c}$ to values obtained with MC events for 
the charmless and charm fit components and allow it to vary for  the \qqbar\ component.

The PDF ${ \cal P}_{\rm sig} (\dt,\, \sigdt; c)$, for each category  $c$, is 
the  convolution of $F(\dt;\, c)$ (Eq.\ \ref{eq:thTime}) with the
signal resolution function (sum of three Gaussians) determined from the
\bflav\ sample. The $\Delta t$ resolution functions for all the other fit 
components are also modeled with  the sum of three Gaussians. For charmless, 
 \Nbtoadueppim, and  $\rho\pi\pi$ components in the nominal fit to the data we assume $S=0$, 
$C=0$, $\Delta S=0$, and $\Delta C =0$  and we vary these parameters   when evaluating 
systematic uncertainties on final results.  We use an effective
$B$ lifetime for the charmless component as obtained from a fit to MC signal events.
The continuum (charm) $\Delta t$ distributions  are parameterized as sums of three Gaussians 
with parameters determined from a fit to off-resonance (MC) events.

The PDF of the invariant mass of the \aunob\ meson in  signal events is parameterized as a 
relativistic Breit-Wigner line-shape with a mass-dependent width that takes into account the
 effect of the mass-dependent $\rho$ width~\cite{WA76}. The PDF of the invariant mass of the 
 ${a_2(1320)}$ meson is parameterized by a sum of three Gaussian function distributions.
The \mes\ and \DE\ distributions for  signal are parameterized as
a sum of two  Gaussian distributions. The \DE\ distribution for continuum
background is parameterized by a linear function, and 
the combinatorial background in \mes\ is described by a phase-space-motivated
empirical function \cite{argus}. We model the
Fisher distribution \xf\ using a Gaussian function with different widths above
and below the mean. The ${\mathcal A}$ distributions are modeled using 
polynomials. 

The PDF parameters are determined from MC simulated events with the exception
of the continuum background, where we use
off-resonance data, and of the signal resolution function, where we use the \bflav\
sample.
Large data control samples of $B$ decays to charmed final states of similar
topology are used to verify the simulated resolutions in \mes\ and \DE.
Where the control samples reveal differences between data and
 MC in mass and energy
resolution, we shift or scale the resolution used
in the likelihood fits. 

We test and calibrate the fitting procedure by applying it to
ensembles of simulated \qqbar\ experiments drawn from the PDF, into which
we have embedded the expected number of signal, charmless, 
 \Nbtoadueppim, the charm, and the $\rho \pi \pi $ 
events randomly extracted from the fully simulated MC samples. The
measured quantities  $\Sa1pi$, $\Ca1pi$,
$\dSa1pi$, $\dCa1pi$, and $\Acpapi$
have been corrected for the fit biases and a systematic uncertainty
equal to half of the bias found in MC simulations is assigned  on the final results. 

In the fit there are 35 free parameters, including $\Sa1pi$, $\Ca1pi$,
$\dSa1pi$, $\dCa1pi$, the  charge asymmetries  for signal and 
continuum background, five yields, 
the signal $\aunob$ width, eleven parameters determining the shape of the combinatorial 
background, and 12 tagging efficiencies for the continuum.
\begin{table}[h]
\caption{Summary of the systematic uncertainties (in units of $10^{-2}$).}
{\small
\begin{center}
\setlength{\tabcolsep}{0.165pc}
\begin{tabular}{lccccc}
\hline\hline
\noalign{\vskip1pt}
 & $S_{\a1pi}$  & $C_{\a1pi}$  &  $\dS_{\a1pi}$  & $\dC_{\a1pi}$  & $\Acpapi$ \\
\noalign{\vskip1pt}
\hline
PDF parameterization             & $4.8$ & $5.3$ & $3.3$ & $5.3$ &  $1.5$  \\
Signal \deltat\ model& $0.2$ & $0.2$ & $0.3$ &$0.1$  &$0.0$    \\
Tagging and mistag   & $0.3$ & $0.2$ & $0.2$ & $0.4$ &  $0.1$  \\
$\deltamd$ and $\tau$ &$0.0$  & $0.2$ & $0.3$ & $0.1$ &$0.0$   \\
Fit bias         &$0.8$  & $0.2$ & $0.8$ & $1.0$ & $0.3$  \\
\BB\ \CP violation & $4.1$ & $4.3$ & $4.2$ &$4.0$  &$0.5$    \\
\Nadueppim\ + interf.    & $2.8$ & $4.5$ & $3.2$ &$0.6$  &$0.2$    \\
$\rm DCS$ decays         &$0.8$  &$2.2$  & $0.0$ & $2.2$ & $0.1$  \\
SVT alignment     & $1.0$ &$0.6$  &$1.0$  &$0.6$  &$0.0$   \\
Particle ID           &$0.1$  &$0.1$  &$0.1$  &$0.1$  & $0.2$  \\
\hline
Total  &$7.0$  & $8.5$ & $6.4$ & $7.1$ & $1.6$  \\
\hline
\hline
\end{tabular}
\end{center}
}
\label{tab:sys_table}
\end{table}
The main contributions to the systematic error on the signal 
parameters are summarized in Table~\ref{tab:sys_table}. 
We have studied  systematic uncertainties arising  from several
sources: variation of the signal PDF shape parameters within their
errors; modeling of the signal \deltat\ distribution;   tagging efficiency 
and mistag rates  determined from
the \bflav\ sample \cite{Resol}; uncertainties in 
$\deltamd$ and $\tau$ \cite{PDG2006}; uncertainty in fit bias; 
uncertainty due to \CP\ violation present in the \BB\ background, the \Nadueppim\
\CP\ violation; uncertainty due to the interference between $B^0  \ra a_1^{\pm} \pi^{\mp}$ and other $4\pi$ final states have been estimated with MC simulations;
doubly-Cabibbo-suppressed (DCS)  $b \ra \bar{u} c
\bar{d}$ amplitude for some tag-side $B$ decays~\cite{Long}; SVT
alignment; and the particle identification algorithm. We allow for a \CP asymmetry 
up to 20\% in  $B$ decays to charmless final states, and up to 
50\% in $B$ decays to ${a_2(1320)} \pi$.
 
From the fit to a sample of 29300 events, we obtain a signal yield of
$608 \pm 53$, of which $461 \pm 46$ have their flavor identified and are used to 
measure the following additional  parameters: 
$\Sa1pi   = 0.37 \pm0.21 \pm 0.07$, $\dSa1pi = -0.14 \pm0.21 \pm0.06 $, 
$\Ca1pi = -0.10\pm 0.15\pm0.09$, $\dCa1pi= 0.26 \pm 0.15\pm
0.07$, $\Acpapi   = -0.07 \pm 0.07 \pm0.02$. Linear correlations
between these fit parameters are small.

The angle  $\alpha_{\rm eff}$ can be defined  \cite{Zupan} as:
\beqn
\label{eq:thTimee}
\alpha_{\rm eff} = \frac{1}{4}
\bigg[\arcsin \bigg(
\frac{\Sa1pi + \dSa1pi }
{\sqrt{1-(\Ca1pi + \dCa1pi)^2}}
\bigg) + \\
	&&\hspace{-4.7cm}\arcsin \bigg(
\frac{\Sa1pi - \dSa1pi }
{\sqrt{1-(\Ca1pi - \dCa1pi)^2}}
\bigg)\bigg]\;\nonumber
\eeqn
Using the measured \CP parameters, we determine the angle $\alpha_{\rm eff}$ and  one of the four solutions,
$\alpha_{\rm eff} = 78.6^{\circ} \pm 7.3^{\circ}$, is compatible with
the result of SM-based fits. Using the published branching
fraction \cite{a1pi} and adding statistical
and systematic errors in quadrature, we obtain also the   
following values for the flavor-charge branching
fractions \cite{Charles} (in units of  $10^{-6}$): ${\cal B}(B^0\ra a_1^+ \pi^-)=17.9\pm4.8$,
${\cal B}(B^0\ra a_1^- \pi^+)=11.4\pm4.7$,
${\cal B}(\Bbar^0\ra a_1^+ \pi^-)=13.0\pm4.3$,
and ${\cal B}(\Bbar^0\ra a_1^- \pi^+)=24.2\pm5.8$.

\begin{figure}[!t]
\resizebox{\columnwidth}{!}{
\begin{tabular}{cc}
\includegraphics[scale=0.25]{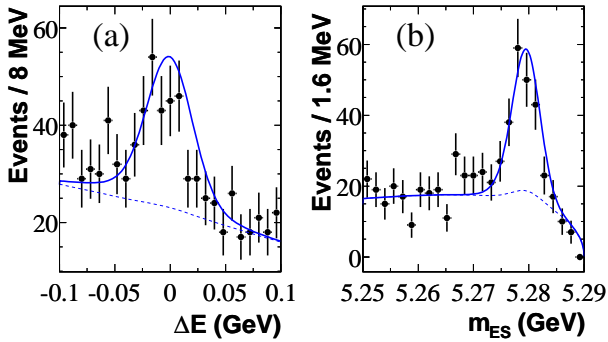} 
\end{tabular}
}
\vspace*{-0.5cm}
\caption{Projections of a) \DE, b)  \mes. 
Points represent on-resonance data, dotted lines 
the sum of all backgrounds, and solid lines the full fit
function. These plots are made with a  cut on the signal likelihood.}
  \label{fig:ProjMesDE}
\end{figure}

\begin{figure}[h]
\includegraphics[scale=0.35]{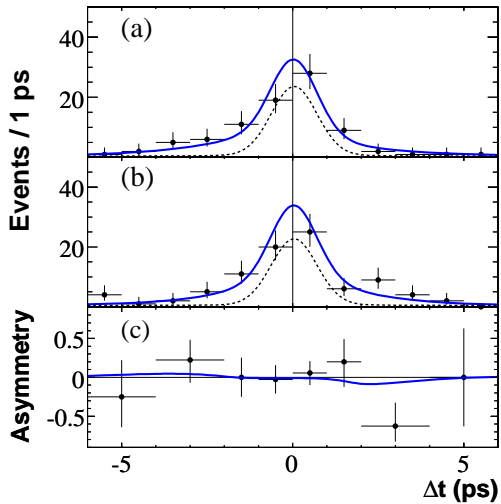} 
\vspace*{-0.5cm}
 \caption{Projections onto $\Delta t$  
 of the data (points) for a) \Bz\ 
and b) \Bzb\  tags, showing the fit function (solid 
line), and the background
function (dotted  line), and c) the asymmetry between \Bz\ and \Bzb\ 
tags.}
  \label{fig:DeltaTProj}
\end{figure}

Figure~\ref{fig:ProjMesDE} shows distributions of $\mes$ and $\DE$, 
enhanced in signal content by requirements on the signal-to-continuum 
likelihood ratios using all discriminating variables other than the
one plotted. 
Figure~\ref{fig:DeltaTProj} gives the $\Delta t$
projections and asymmetry for flavor tagged events.

In summary, we have measured the \CP-violating
asymmetries in  
\Nbtoappim\ decays and determined the angle $\alpha_{\rm eff}$. We do not find 
evidence for direct or mixing-induced \CP violation in these
decays. Once 
measurements of branching fractions for SU(3)-related decays become
available, quantitative bounds on $\Delta \alpha$ obtained with the
method
of Ref. \cite{Zupan} will provide significant constraints on the angle $\alpha$
through the measurement of $\alpha_{\rm eff}$ in \Nbtoappim\ decays.

We are grateful for the excellent luminosity and machine conditions
provided by our \pep2\ colleagues, 
and for the substantial dedicated effort from
the computing organizations that support \babar.
The collaborating institutions wish to thank 
SLAC for its support and kind hospitality. 
This work is supported by
DOE
and NSF (USA),
NSERC (Canada),
IHEP (China),
CEA and
CNRS-IN2P3
(France),
BMBF and DFG
(Germany),
INFN (Italy),
FOM (The Netherlands),
NFR (Norway),
MIST (Russia), and
PPARC (United Kingdom). 
Individuals have received support from CONACyT (Mexico), 
Marie Curie EIF (European Union),
the A.~P.~Sloan Foundation, 
the Research Corporation,
and the Alexander von Humboldt Foundation.


\begin{thebibliography}{99}
\bibitem{a1}
For the $a_1(1260)$ meson we use the short notation ${a_1}$.
\end{thebibliography}
\end{document}